\documentclass{kapproc}
\usepackage[latin1]{inputenc}
\usepackage{amsmath}
\usepackage{amssymb}
\usepackage[dvips]{graphicx}
\upperandlowercase
\normallatexbib %will produce bibliography entries as shown in the
\begin{document}

\articletitle{Shot Noise in Mesoscopic Diffusive Andreev Wires}

\articlesubtitle{}

\author{Wolfgang Belzig}
\affil{Department of Physics and Astronomy, University of Basel,\\
Klingelbergstr. 82, CH-4056 Basel, Switzerland}
\email{Wolfgang.Belzig@unibas.ch}

%\author{Second Author}
%\affil{Author Affiliation\\
%Second Line of Affiliation}
%\email{secondauthor@anotheruniv.edu}

\begin{abstract}
  We study shot noise in mesoscopic diffusive wires between a normal
  and a superconducting terminal. We particularly focus on the regime,
  in which the proximity-induced reentrance effect is important. We
  will examine the difference between a simple Boltzmann-Langevin
  description, which neglects induced correlations beyond the simple
  conductivity correction, and a full quantum calculation. In the
  latter approach, it turns out that two Andreev pairs propagating
  coherently into the normal metal are anti-correlated for $E\lesssim
  E_c$, where $E_c=\hbar D/L^2$ is the Thouless energy. In a fork
  geometry the flux-sensitive suppression of the effective charge was
  confirmed experimentally.
\end{abstract}


\section{Introduction}

Fluctuations of the current in mesoscopic conductors originate from
the quantum scattering of the charge carriers, and are sensitive to
their interference, statistics and interaction. This makes the theoretical
and experimental study of noise in small electronic circuits
interesting and challenging (for recent reviews, see
Refs.~\cite{blanter,nazarov:03}).

Statistical correlations in the transport of fermions have led to a
number of interesting predictions.  For example, the noise of a
single-channel quantum contact of transparency $T$ at zero temperature
has the form $S_I=2|eI|(1-T)$
\cite{khlus:87,lesovik:89,buettiker:90,buettiker:92}. The noise is
thus suppressed below the Schottky value $2|eI|$ of uncorrelated
charge transfer. The suppression is a direct consequence of the Pauli
principle, which prevents two electrons from tunneling together.
It is convenient to measure the deviation from the Schottky result by
the so-called Fano factor $F=S_I/2|eI|$. For a number of generic
conductors, it turns out that the suppression of the Fano factor is
universal, i.~e. it does not depend on details of the conductor like
geometry or impurity concentration. In particular, a diffusive metal
with elastic scattering leads to $F_{diff}^N=1/3$
\cite{been-buett:92,nagaev:92}, which is independent on the concrete
shape of the conductor \cite{yuli:94-diffusive} and has been confirmed
experimentally \cite{dieleman:97,hoss:00}.

If superconductivity comes into play the fundamental charge transport
mechanism at energies below the superconducting gap is Andreev
reflection. Two electrons enter the superconductor simultaneously and
form a Cooper pair, which can propagate in the superconductor.  Thus,
in this process a charge transfer of $2e$ occurs, but with a reduced
probability, since two particles have to tunnel. The shot noise is
proportional to the charge of the elementary processes, and one thus
naivly expects a doubling of the shot noise, which was indeed found
theoretically \cite{jong:94,nagaev:01} and experimentally
\cite{jehl:99,kozhevnikov:00} for diffusive conductors.  It is
remarkable that this doubling occurs for diffusive conductors, whereas
it is not found for other conductors like, e.~g., single-channel
contacts \cite{khlus:87,jong:94,khmelnitzkii:94,martin:96,anantram:96}
, double tunnel junctions
\cite{naidenov:95,fauchere:98,schep:97,samuelsson:03}, or diffusive
junctions with a tunneling barrier \cite{stenberg:02,pistolesi:03}. A
doubling of the full Schottky noise was recently observed
experimentally \cite{lefloch:03}.

In this article we address the counting statistics and the noise in
diffusive structures with normal and/or superconducting terminals. In
particular we concentrate on the energy- and phase-dependent shot
noise in an Andreev interferometer. First, we briefly review the
theory of full counting statistics using the Keldysh Green's function
approach. We derive the counting statistics of diffusive conductors
for various limits. In Sec.~3 we obtain generic results for the shot
noise in diffusive conductors in some limiting cases.  Finally, in
Secs.~4 and 5 we discuss the information contained in the energy- and
phase-dependence of the shot noise in diffusive wires and Andreev
interferometers. A good qualitative agreement of experimental results
and our full quantum calculation is demonstrated.

\section{Current Statistics}
\label{sec:general}

Consider the following Gedanken experiment. A constant voltage bias is
applied to a mesoscopic conductor for a certain time interval $t_0$.
During this time interval we count the number of charges $N$ passing
the conductor. Due to quantum-mechanical uncertainty the outcome of
the experiment is described by a probability $P_{t_0}(N)$ that $N$
charges have passed the conductor. This is the so-called \textit{full
  counting statistics} (FCS), which is the most natural
description of quantum transport. Alternatively we may study the
\textit{characteristic function} $\Phi(\chi)=\sum_N P_{t_0}(N)
e^{iN\chi}$. The $k$th coefficient in the expansion of the
characteristic function in powers of $\chi$ yields the moments of the
counted charge $\langle N^k \rangle$. An equivalent description is to
obtain the \textit{cumulant generating function} (CGF)
$S(\chi)=\ln\Phi(\chi)$, which gives the cumulants directly. This will
be the quantity which we obtain below. The cumulants are directly
connected to the (measureable) zero-frequency current correlations
functions. 

One route to counting statistics is the Keldysh-Green's function
approach in combination with the circuit theory of mesoscopic
transport developed by Nazarov
\cite{yuli:94-circuit,yuli:99-annals,yuli:99-supplat}. For details, we
refer to several articles in \cite{nazarov:03}. In this approach,
terminals are described by 4$\times$4 Green's function matrices. For a
normal terminal (N) with occupation $f$ we introduce the 2$\times$2
matrix $\hat
f=$diag$(f(E),f(-E))$ and have
\begin{equation}
  \check G_N(\chi)=\left(
    \begin{array}[c]{cc}
      \hat\tau_3(1-2\hat f) & 
      -2\hat\tau_3 e^{i\chi\hat\tau_3} \hat f\\
      -2\hat\tau_3 e^{-i\chi\hat\tau_3} \hat f &
      \hat\tau_3(2 \hat f-1) 
    \end{array}\right)\,.
%  \,,\,
%  \check G_N(E,\chi)=e^{i\chi\check \tau_K/2} \check G_N(E) e^{-i\chi\check \tau_K/2}
\end{equation}
For a superconducting terminal (S) at equilibrium and for $E\ll\Delta$
we have
\begin{equation}
  \check G_S = \left(
    \begin{array}[c]{cc}
      \hat\tau_1 & 0 \\
      0 & \hat \tau_1
    \end{array}\right)\,.
\end{equation}
For a general contact described by a scattering matrix with
corresponding transmission eigenvaluese  $\{T_n\}$ the counting
statistics is obtained as \cite{belzig:01-super}
\begin{equation}
  \label{eq:fcs-general-scatterer}
  S(\chi)=\frac{t}{h}\int dE \sum_n \textrm{Tr} 
  \ln\left[1+\frac{T_n}{4}\left(\left\{\check G_L,\check G_R\right\}-2\right)\right]\,.
\end{equation}
To give a simple example, we evaluate the counting statistics of a
quantum contact between two normal terminals and obtain in agreement
with the scattering matrix approach
\cite{levitov:93-fcs,levitov:96-coherent} (see Appendix for a short summary)
\begin{equation}
  \label{eq:fcs1}
  S(\chi)=\frac{t}{h}\int dE \sum_n 
  \ln\Big[1
    \begin{array}[t]{l}
      +\;T_nf_L(1-f_R)(e^{i\chi}-1)\\
      +\;T_nf_R(1-f_L)(e^{-i\chi}-1)\Big]\,.
  \end{array}
\end{equation}
Thus, the statistics is a simple multinomial form of the two possible
events of left and right transfer of charges.  The formula
(\ref{eq:fcs-general-scatterer}) includes in addition the statistics
of SN- and SS-contacts \cite{khmelnitzkii:94,belzig:01-super}.

\begin{figure}[tbp]
  \centering
  \includegraphics[width=\textwidth]{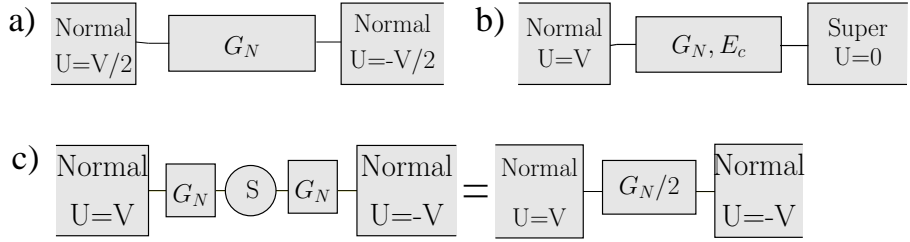}
  \caption{a) diffusive wire with conductance $G_N$ between two normal terminals. b)
    diffusive wire betweena normal and a superconducting terminal. The
    wire is characterized by the characteristic energy $E_c=\hbar
    D/L^2$. c) in the incoherent regime the wire b) is mapped onto a
    series of two wires between normal terminals. For diffusive
    connectors the system is equivalent to one wire with conductance
    $G_N/2$.}
  \label{fig:system}
\end{figure}

In diffusive conductors we have to find the general counting
statistics by a different method. The quantum kinetic equation for the
diffusive wire is the so-called Usadel equation \cite{usadel}
\begin{equation}
  \label{eq:usadel}
   \hbar\frac{\partial}{\partial x} D(x) \check G(x) 
  \frac{\partial}{\partial x} \check G(x) =
  -i\left[E\hat\tau_3,\check G(x)\right]\,,
\end{equation}
with continuous boundary conditions at both ends of the conductor. The
right-hand side of Eq.~(\ref{eq:usadel}) accounts for the decoherence
of electrons and holes during the propagation in the normal metal at
finite energies $E$. A full solution has so far been only obtained
numerically, and we will discuss the implications of the decoherence
on the shot noise later.

For diffusive wires between normal terminals or with one
superconducting terminal (see Fig.~\ref{fig:system}a and
Fig.~\ref{fig:system}b for smalle energies $eV,k_BT \ll E_c$) the
right hand side of Eq.~(\ref{eq:usadel}) can be neglected. We obtain the
general solution
\cite{yuli:99-annals,levitov:96-diffusive,belzig:03-book}
\begin{equation}
  \label{eq:fcs-diff-general}
  S(\chi)=\frac{t_0G_N}{8h}\int dE\,\textrm{Tr}\, 
  \textrm{acosh}^2\left(\frac 12\left\{\check G_L,\check G_R\right\}\right)\,.
\end{equation}
The same result can be obtained by averaging
Eq.~(\ref{eq:fcs-general-scatterer}) over the transmission eigenvalue
distribution of a diffusive scatterer, i.e.  the bimodal distribution
\cite{dorokhov}
\begin{equation}
  \rho(T)=\frac{G_N}{2G_Q}\frac{1}{T\sqrt{1-T}}\,.
  \label{eq:dorokhov}
\end{equation}

Another drastic simplification can be made if the proximity effect is
negligible , i.~e. the right hand side of
Eq.~(\ref{eq:usadel}) is dominant. In Ref.~\cite{belzig:03} it was
shown, that the diffusive wire can be mapped
onto a series of two diffusive wires contacted by normal terminals,
which constitute the electron and hole propagation (see
Fig.~\ref{fig:system}c). For the special case of diffusive connectors
the counting statistics is independent of the geometry. As a
consequence, the counting statistics is exactly given by that of a
\textit{normal} conductor, with \textit{halved} conductance and a
\textit{negative} counting field for the hole terminal.

The energy dependence resulting from the right hand side of
Eq.~(\ref{eq:usadel}) leads to interesting effects related to the
quantum propagation of electron-hole pairs \cite{belzig:01-diff}.
Below, we will address in detail the dependence of the shot noise on
voltage and temperature, and compare our theoretical predictions,
based on (\ref{eq:usadel}) with available experimental results.

\section{Shot Noise in Diffusive Conductors}
\label{sec:noise}

To obtain the shot noise from the counting statistics we calculate
\begin{equation}
  S_I=-\frac{2e^2}{t_0}
  \left.\frac{\partial^2}{\partial\chi^2} S(\chi)\right|_{\chi=0}\,.
\end{equation}
From Eq.~(\ref{eq:fcs-general-scatterer}) we obtain for a general
scatterer between normal terminals
\begin{eqnarray}
  \label{eq:si-diff-general}
  S_I^{NN} & = & \frac{2e^2}{h} \sum_n \int dE 
  \bigg[T_n(1-T_n)\left(f_L(E)-f_R(E)\right)^2\\\nonumber
  && +T_n \big[f_L(E)(1-f_L(E))+f_R(E)(1-f_R(E))\big]\bigg]\,.
\end{eqnarray}
Averaging (\ref{eq:si-diff-general}) over the transmission eigenvalue
distribution (\ref{eq:dorokhov}) we find
\begin{equation}
  \label{eq:noise-diff-normal}
  S_I^{NN}  =  2 G_N \int dE 
  \begin{array}[t]{r}\displaystyle
    \bigg[f_L(E)\left(1-f_L(E)\right)+f_R(E)\left(1-f_R(E)\right)\\\displaystyle
    +\frac13\left(f_L(E)-f_R(E)\right)^2\bigg]\,.
  \end{array}
\end{equation}
The energy integration can be done using the formulas in the
appendix. As result we obtain
\begin{equation}
  \label{eq:noise-diff-scatter}
  S_I^{NN}(T,V)=\frac{4}{3} G_N k_BT + 
  \frac{2}{3}eG_NV \textrm{coth}\left(\frac{eV}{2k_BT}\right)\,,
\end{equation}
where we introduced the conductance $G_N=\frac{e^2}{h}\sum_n T_n$.

Using the result (\ref{eq:fcs-diff-general}) we can obtain at $E=0$
the general formula for the noise
\begin{equation}
  \label{eq:noise-diff-general}
  S_I=-\frac{G}{4h}\int dE\,\textrm{Tr}\,\left[ 
  \frac{\partial^2}{\partial\chi^2}\left\{\check G_L,\check G_R\right\}
  -\frac{2}{3}\left(\frac{\partial}{\partial\chi}
      \left\{\check G_L,\check G_R\right\}\right)^2\right]_{\chi\to 0}\,.
\end{equation}
%% This result can be further reduced using
%% \begin{equation}
%%   -\frac{\partial}{\partial\chi}\check G(\chi)=
%%   \frac12\left[\check\tau_K,\check G(\chi)\right]\,.
%% \end{equation}
%% For two normal terminals we find the result
%% (\ref{eq:noise-diff-scatter}).
For a diffusive wire between a normal and a superconducting terminal
at $eV,k_BT\ll E_c$ we obtain
\begin{equation}
  S_I^{NS}(V)=\frac{2G_N}{3}\int dE \left[f(E)+f(-E)\right]\left[2-f(E)-f(-E)\right]\,.
\end{equation}
Evaluating the energy integration we find
\begin{equation}
  \label{eq:si-diff-sn}
  S_I^{NS}(V)=\frac{8}{3}G_Nk_B T + 
  \frac{4}{3}eG_N V \textrm{coth}\left(\frac{eV}{k_BT}\right)\,.
\end{equation}
This result can, indeed, be infered from the normal-state result
Eq.~(\ref{eq:noise-diff-general}) by the replacement $G_N\to 2G_N$ and
$e\to 2e$.

While the derivations presented previously are valid in the limit
$eV$, $k_BT$ $\ll$ $E_c$, we can also obtain the noise in the limit $eV\gg
E_c$ or $k_BT\gg E_c$. Here we employ the incoherent Andreev circuit
theory approach \cite{belzig:03}. According to the mapping rules we
obtain the noise from the normal result by the replacement $G_N\to
G_N/2$, $V\to 2V$ and $S_I\to 4S_I$. By applying these substitutions
to Eq.~(\ref{eq:noise-diff-scatter}) we again obtain the result
(\ref{eq:si-diff-sn}).

We see that the shot noise (as well as the full counting statistics)
in the discussed regimes is universal. First, at low energies
$eV,k_BT\ll E_c$ universality means, that the noise depends only
on the normal-state conductance $G_N$ and is independent of the
detailed geometry. Furthermore, it turns out that in the incoherent
regime the full counting statistics and, therefore the current noise
is also the same. The universality of the noise is quite surprising,
since the transport mechanisms differ quite drastically in both
limits.  This remarkable coincidence holds, however, only for
diffusive conductors. For double tunnel junctions, chaotic cavities,
or other combinations of scatterers the transport properties differ
(see Ref.~\cite{belzig:03-book} and references therein).

\section{Energy-dependent current noise}
\label{sec:energy}

\begin{figure}[th]
  \centering
  \begin{minipage}[b]{0.3\textwidth}
    \includegraphics[width=\textwidth]{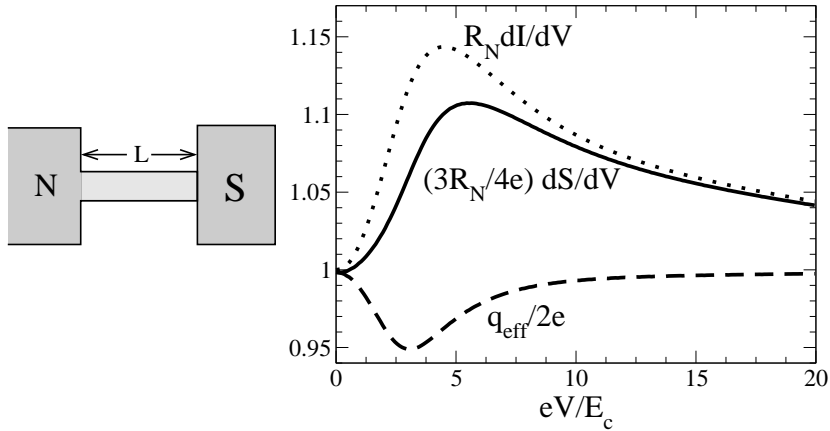}\\[2cm]
  \end{minipage}
  \includegraphics[width=0.6\textwidth,clip]{wire_noise.eps}
  \caption{Shot noise of a diffusive proximity wire. Both the
  differential conductance and the noise show a reentrant
  behaviour. The effective charge reveals that the correlated Andreev
  pair transport suppresses the noise below the uncorrelated
  Boltzmann-Langevin result.}
  \label{fig:wire}
\end{figure}

The full quantum-mechanical description requires the solution of the
Usadel equation (\ref{eq:usadel}) to first order in the counting
field. We will discuss these results later.  Let us first note, that
one can obtain an approximate expression for the energy dependence of
the shot noise by a generalized Boltzmann-Langevin approach. We
recall, that the kinetic equation for the average distribution
function has the form
\begin{equation}
  \label{eq:usadel-kinetic}
  \frac{\partial}{\partial x}\sigma(E,x) 
  \frac{\partial}{\partial x} \left(1-f(E,x)-f(-E,x)\right) = 0\,.
\end{equation}
The local energy-dependent conductivity incorporates the effect of the
proximity induced coherence and is given by $\sigma(E,x) =
\sigma_N$ $\cosh(\textrm{Re}\theta(E,x))$, where the spectral angle
$\theta$ obeys $\hbar D(\partial^2/\partial
x^2)\theta(E,x)=-iE\sin(\theta(E,x))$ with appropriate boundary
conditions (see Ref.~\cite{yuli:96} for details).  Solving
Eq.~(\ref{eq:usadel-kinetic}) the current (per unit area of the
contact) is
\begin{equation}
  \label{eq:usadel-current}
  I(V,T)=\frac{1}{e}\int dE G(E) \left[1-f_N(E)-f_N(-E)\right]\,.
\end{equation}
Here we have defined the spectral conductance
\begin{equation}
  \label{eq:spectral-conductance}
  \frac{1}{G(E)}=\int_0^L \frac{dx}{L} \frac{1}{\sigma(E,x)}\,,
\end{equation}
which have written here for a one-dimensional wire of length $L$ with a uniform
cross section. For an arbitrary geometry, the spectral conductance has
to be found from the solution of a diffusion equation.

Guided by the kinetic equation for the average transport, we may try
to find the current noise by generalizing the Boltzmann-Langevin
approach \cite{nagaev:01} to include the energy- and space-dependent
conductivity. In fact, this task has already been performed. In
Ref.~\cite{yuli:99-annals} it was shown, that the counting statistics
of a diffusive normal wire only depends on the conductance, defined in
exactly the same way as (\ref{eq:spectral-conductance}). In this proof
an energy-independent conductivity $\sigma(x)$, but an arbitrary form
of the diffusive metal was taken into account.  However, it is
obvious, that the same result is obtained if a spectral conductivity
$\sigma(E,x)$ is assumed. As a result the current would be given by
Eq.~(\ref{eq:usadel-current}).

Next, we combine this observation with the results for incoherent
Andreev transport obtained in Ref.~\cite{belzig:03}. In this work it
was shown that the counting statistics of a diffusive wire between a
normal and a superconducting terminal is mapped onto a series of two
diffusive wires. Again, this holds equally well if we assume
\textit{ad hoc} an energy-dependent conductance of both diffusive
wires. Due to electron-hole symmetry the spectral conductance of both
wires and therefore of the total wire is the same.

Collecting these observations, we obtain the current noise from
Eq.~(\ref{eq:noise-diff-normal}) by inserting the spectral conductance
(\ref{eq:spectral-conductance}) \textit{inside} the energy integral
and multiplying by a factor of 2. In this way we obtain the result of
the modified Boltzmann-Langevin approach
\begin{equation}
  S_I^{BL}  =  4 \int dE G(E) 
  \begin{array}[t]{r}
    \Big[f(E)(1-f(E))+f(-E)(1-f(-E))\\+\frac13(f(E)-f(-E))^2  \Big]
  \end{array}\,.
\end{equation}
Evaluating the energy integral with the help of the formulas in the
appendix we find
\begin{equation}
  \label{eq:noise-usadel}
  S_I^{BL} = \frac{8}{3} G(V,T) k_B T 
  + \frac{4}{3} e I(V,T) \textrm{coth}\left(\frac{eV}{k_BT}\right)\,.
\end{equation}
Here we introduced the temperature-dependent differential conductance
$G(V,T)=dI(V,T)/dV$. The same result was obtained by a different
method recently in Ref.~\cite{houzet:03}. At zero temperature, we
obtain the result \cite{reulet:03-book}
\begin{equation}
  \label{eq:noise-bl}
  S_I^{BL}(V)=\frac{4}{3}eI(V)\,.
\end{equation}
This result represents the starting point for our further
considerations. It was derived neglecting correlations of scattering
events \textit{between electrons and holes} in the normal metal. Thus,
in the following we will be specifically interested in the
deviations of the noise from the simple Boltzmann-Langevin result
(\ref{eq:noise-bl}) and introduce the \textit{effective charge}
$q_{eff}(V)=(3/2)\partial S_I/\partial I$ \cite{reulet:03}.

The full quantum-mechanical calculation of the energy-dependent shot
noise was performed in Ref.~\cite{belzig:01-diff} and the results are
shown in Fig.~\ref{fig:wire}. A direct comparison of the differential
shot noise and the differential conductance (for zero temperature)
shows the difference in the energy dependence. The
effective charge defined above displays the clear deviation of the quantum noise
from the Boltzmann-Langevin result of $2e$. At energies below the
Thouless energy $E_c=\hbar D/L^2$ the effective charge is suppressed
below $2e$. This shows that the correlated Andreev pair transport
suppresses the noise below the uncorrelated Boltzmann-Langevin result.

\begin{figure}[th]
  \centering
  \begin{minipage}[b]{0.3\textwidth}
    \includegraphics[width=\textwidth]{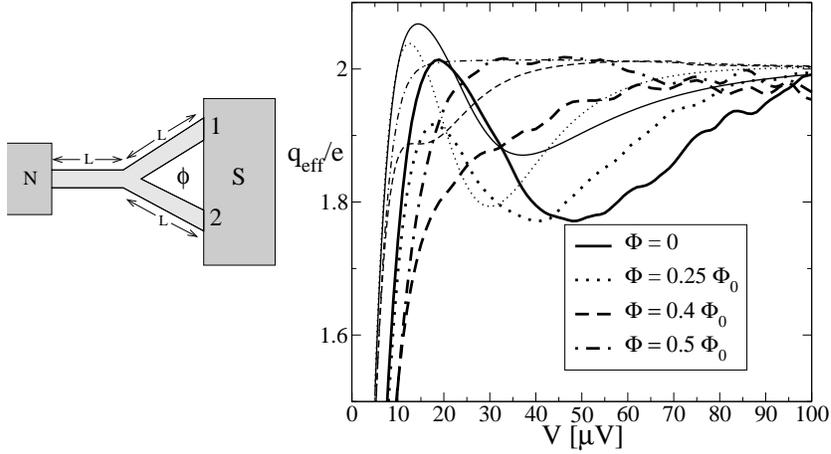}\\[2cm]
\end{minipage}
  \includegraphics[width=0.6\textwidth]{fork_noise_experiment.eps}
  \caption{Effective charge of the Andreev interferometer shown in the
    left (realized experimentally in Ref.~\cite{reulet:03}).  The
    right panel shows the experimental (thick lines) and theoretical
    (thin lines) results for the effective charge are shown in the
    main plot for different magnetic fluxes. The theoretical results
    contain no fitting parameter, since the only relevant energy scale
    $E_c=30\mu$eV was extracted from the sample geometry. Therefore,
    we believe the deviations between experimental and theoretical
    results comes from possible heating effects in the experiment,
    which are not accounted for in the theoretical calculation.}
  \label{fig:fork}
\end{figure}

\section{Phase-dependent shot noise}
\label{sec:phase}

To experimentally probe the pair correlations in diffusive
super\-con\-duc\-tor-normal metal-hetero\-structures it is most
convenient to use an Andreev interferometer. An example is shown in
the left part of Fig.~\ref{fig:fork}. A diffusive wire connected to a
normal terminal is split into two parts, which are connected to two
different point of a superconducting terminal. By passing a magnetic
flux through the loop one can effectively vary the phase difference
between the two connections to the superconductor. Such a structure
has been experimentally realized by the Yale group \cite{reulet:03}.
In Fig.~\ref{fig:fork} we present a direct comparison between our
theoretical predictions and the experimentally obtained effective
charge. Note that we have included the experimental temperature in the
theoretical modelling.  The finite temperature explains the strong
decrease of the effective charge in the regime $|eV|\leq k_BT$, where
the noise is fixed by the fluctuation-dissipation theorem. The
disagreement between theory and experiment in this regime stems solely
from differences in the measured temperature-dependent conductance
from the theoretical prediction. We attribute this to heating effects.
The qualitative agreement in the shot-noise regime $|eV|\geq k_BT$ is
satisfactory, if one takes into account, that we have no free
parameters for the theoretical calculation. Both, experiment and
theory show a suppression of the effective charge for some finite
energy, which is of the order of the Thouless energy and depends on
flux in a qualitative similar manner. Remarkably for half-integer flux
the effective charge is completely flat, in contrast to what one would
expect from circuit arguments based on the conductance distribution in
the fork geometry. Currently we have no explanation for this
behaviour, and therefore more work is needed in this direction.

\section{Conclusions}
\label{sec:concl}

Shot noise in diffusive heterostructures between normal and
superconducting terminal provides valuable information on the
correlated Andreev pair transport. We have examined the difference
between a simple Boltzmann-Langevin description, which neglects these
correlations, and a full quantum calculation. Examining the effective
charge we have shown that two Andreev pairs propagating coherently
into the normal metal are anti-correlated. The phase sensitivity of
the suppression of the effective charge was confirmed experimentally.

We thank C. Bruder, Yu. V. Nazarov, D. Prober, B. Reulet, and
P. Samuelsson for discussions. This work was supported by the Swiss
National Science Foundation and the NCCR Nanoscience.

\chapappendix{Scattering Approach}

One of the most general approaches to quantum transport is the
scattering matrix approach. Following Levitov and Lesovik
\cite{levitov:93-fcs,levitov:96-coherent} the counting statistics can
be found from the a slight modification of the standard scattering
matrix approach. For a two terminal structure we label the scattering
states in the leads by
$\psi=(c_1^L,\ldots,c_{n_L}^L,c_1^R,\ldots,c_{n_R}^R)$. The modified
scattering matrix and the occupation matrix read
\begin{equation}
  S_\chi=\left(
    \begin{array}[c]{cc}
      r  & t e^{i\chi/2} \\
      t^\prime e^{-i\chi/2} & r^\prime
    \end{array}\right)\quad;\quad
  f=\left(
    \begin{array}[c]{cc}
      f_L  & 0 \\
      0 & f_R
    \end{array}\right)\,.
\end{equation}
The counting statistics is then given by
\begin{equation}
  \Phi_E(\chi)=\textrm{det}\left(1-f+fS_\chi S_{-\chi}^\dagger\right)
  \,,\, S(\chi)\equiv\ln\Phi(\chi)=\frac{t_0}{h}\int dE \ln \Phi_E(\chi) \,.
\end{equation}
Using the standard polar decomposition for the scattering matrix we
obtain the counting statistics (\ref{eq:noise-diff-general}).

\chapappendix{Integrals}

In the derivations in this article we encounter integral expressions
of the forms
\begin{eqnarray}
  I_1(U) & = & \int dE G(E)\left[f_1(E)(1-f_1(E))+f_2(E)(1-f_2(E))\right]\,,\\
  I_2(U) & = & \int dE G(E)\left[f_1(E)(1-f_2(E))+f_2(E)(1-f_1(E))\right]\,.
\end{eqnarray}
Here we introduced $f_1(E)=f_D(E+U)$ and $f_2=f_D(E-U)$, where
$f_D(E)= (\exp(E/k_BT)+1)^{-1}$ is the Fermi-Dirac distribution.  The
goal is to reduce the integrals to expressions related to the current
$I(U)=\int dE G(E) (f_1(E)-f_2(E))$. The first integral is solved by
noting that $-k_BT(\partial/\partial U) f_{1(2)}(E)=\pm f_{1(2)}(E)(1-f_{1(2)}(E))$
and we find for the first integral
\begin{equation}
  \label{eq:i1}
  I_1(U)= k_BT \frac{\partial I(U)}{\partial U}\,.
\end{equation}
With the help of the identity 
\begin{equation}
  f_1(E)(1-f_2(E))+f_2(E)(1-f_1(E)) =
  (f_1(E)-f_2(E))\coth\left(\frac{U}{k_BT}\right)\,,
  \label{eq:ident}
\end{equation}
we find for the second integral
\begin{equation}
  \label{eq:i2}
  I_2(U)=I(U)\coth\left(\frac{U}{k_BT}\right)\,.
\end{equation}

%% \chapappendix{Derivatives}
%% In the derivation of the noise of a diffusive wire we encounter the derivative
%% \begin{equation}
%%   \begin{array}[c]{ll}\displaystyle
%%     \frac{\partial^2}{\partial\chi^2} \textrm{acosh}^2[f(\chi)] = 
%%     2 \frac{\partial}{\partial\chi}\left[\frac{\partial f(\chi)}{\partial\chi}
%%     \frac{\textrm{acosh}(f(\chi))}{\sqrt{f^2(\chi)-1}}\right]\\\displaystyle
%%       =   \frac{\partial^2f(\chi)}{\partial\chi^2}
%%     \frac{2\textrm{acosh}(f(\chi))}{\sqrt{f^2(\chi)-1}}
%%     -2\left(\frac{\partial f(\chi)}{\partial\chi}\right)^2
%%     \frac{f(\chi)\textrm{acosh}(f(\chi))-\sqrt{f^2(\chi)-1}}{\left(f^2(\chi)-1\right)^{3/2}}
%%   \end{array}
%% \end{equation}
%% In the interesting limit $\chi\to 0$ in our case we have to take the
%% limit $f\to 1$. We obtain, using l`Hospital,
%% \begin{equation}
%%   2\left.\frac{\partial^2f(\chi)}{\partial\chi^2}\right|_{\chi\to 0}
%%   -\frac{2}{3}\left.\left(\frac{\partial
%%   f(\chi)}{\partial\chi}\right)^2\right|_{\chi\to 0} \,,
%% \end{equation}
%% yielding Eq.~(\ref{eq:noise-diff-general}).

\end{document}